\documentclass[reprint,aps,prb,showpacs]{revtex4-1}
\usepackage{graphics}
\usepackage{graphicx,epsfig}
\usepackage{epstopdf}
\usepackage{amsmath}

\begin{document}


\title{Interactions between unidirectional quantized vortex rings} 


\affiliation{School of Physics and Astronomy, The University of Manchester, Oxford Road, Manchester, M13 9PL, United Kingdom}
\author{T. Zhu}
	\altaffiliation{Present address: Department of Physics and Astronomy, Rutgers University, Piscataway, NJ 08854, USA.}

\author{M. L. Evans}	
	\altaffiliation{Present address: Cavendish Laboratory, University of Cambridge, Cambridge CB3 0HE, UK.}

\author{R. A. Brown}	

\author{P. M. Walmsley}	

\author{A. I. Golov}	
	

\date{\today}

\begin{abstract}
We have used the vortex filament method to numerically investigate the interactions between pairs of quantized vortex rings that are initially traveling in the same direction but with their axes offset by a variable impact parameter. The interaction of two circular rings of comparable radii produce outcomes that can be categorized into four regimes, dependent only on the impact parameter; the two rings can either miss each other on the inside or outside, or they can reconnect leading to final states consisting of either one or two deformed rings. The fraction of energy went into ring deformations and the transverse component of velocity of the rings are analyzed for each regime. We find that rings of very similar radius only reconnect for a very narrow range of the impact parameter, much smaller than would be expected from the geometrical cross-section alone. In contrast, when the radii of the rings are very different, the range of impact parameters producing a reconnection is close to the geometrical value. A second type of interaction considered is the collision of circular rings with a highly deformed ring. This type of interaction appears to be a productive mechanism for creating small vortex rings. The simulations are discussed in the context of experiments on colliding vortex rings and quantum turbulence in superfluid helium in the zero temperature limit.

\end{abstract}

\pacs{67.25.dk,47.32.cf,47.27.Cn}

\maketitle 

\section{Introduction}
Vortex rings\cite{shariff92,barenghi09} are a common feature of many different fluid systems and can occur on a huge variety of length scales ranging from nanometers to interstellar plasmas. A circular ring moves forward due to its own self-induced velocity. Vortex rings in superfluids and Bose-Einstein condensates are perhaps the most ideal realization of vortex rings in nature. This is due to the topological stability provided by the quantization of circulation which means that there can be no gradual decay of vorticity as the ring moves. In the case of superfluid helium, the very small fixed core size relative to the ring radii allows any dynamics associated with the core to be neglected. This means that vortex rings in a superfluid near zero temperature, where there is very little dissipation, can travel distances that are many orders of magnitude greater than their radius. In contrast, vortex rings in classical fluids, which usually have relatively thick cores, tend to  break up due to greater dissipation and core instabilities.

There have been many investigations of interacting and reconnecting vortex rings with the post-interaction state consisting of either a single ring or several rings depending on the initial conditions. The cases where two rings are moving towards each other, either at an angle or head-on but with their axes offset, are the ones that have been investigated in most detail in both classical\cite{kambe71,fohl75} and quantum fluids\cite{koplik96,leadbeater01,chatelain03,caplan14}. The scenario where two rings are initially traveling in the same direction has not been studied in as much detail\cite{oshima77,leadbeater03} and this paper seeks to address this issue. In this case, the relative velocity of the rings is much lower, which allows a much longer time for non-local effects to act resulting in several novel outcomes. It was recently suggested that interactions within a turbulent vortex tangle between vortex loops traveling in the same direction can lead to energy being transferred to both larger and shorter length scales\cite{baggaley12b,baggaley14}. This paper focuses on the interaction between isolated pairs of unidirectional rings, our analysis is based on extracting the effective post-interaction radii of the rings and using these to categorize the outcomes of the interactions. The special case of  leapfrogging, that occurs when the rings are coaxial, has already been studied extensively\cite{caplan14,wacks14} and is thus not investigated any further here.

In the helium superfluids, it is possible to create a beam containing a large number of vortex rings traveling in the same direction and with almost identical radii \cite{gamota71,guenin78}. It is now well established experimentally that collisions within such a beam of unidirectional vortex rings\cite{bradley05,walmsley08} can cause the onset of quasi-classical quantum turbulence (tangles of quantized vortex lines with instantaneous correlations of orientation)  which will have fluctuations of coarse-grained velocity over large length scales -- in contrast to uncorrelated tangles of vortices. The interactions between pair of unidirectional rings is highly relevant to this strongly anisotropic vortex state and could shed some light on the underlying microscopic processes such as vortex reconnections and the transfer of energy to deformations of vortex lines (Kelvin waves) which can then induce self-reconnections and the emission of small rings. There have been several simulations that have looked at the build-up of turbulent tangles by colliding many rings together\cite{barenghi02,fujiyama10,baggaley12b,baggaley14} although these had relatively low resolution such that the finer details of ring-ring interactions are not that apparent.

One recent experiment used time-of-flight spectroscopy to observe the effect of reconnections within such a beam\cite{walmsley14}. One of the main observations of this work was the frequent creation of small rings (with radii much smaller than the incident rings, e.g. less than half). The exact mechanism leading to the creation of these small rings is unknown although the production of small rings is a common feature in many theoretical and computational works on quantum turbulence near zero temperature\cite{svistunov95,kerr11,kursa11,tsubota00,kozik09,yepez09,nem10,kond12,salman13,nem14,kond14,laurie15}. Yet the question of whether the interaction between a pair of rings can produce rings of much smaller radii has not yet been addressed, and this paper seeks to shed further light on this issue by examining the efficiency for the creation of small rings when larger rings (both circular and deformed) interact with each other. 

The velocity, $v_0$, energy, $E_0$, and impulse $I_0$, for a circular vortex ring of radius $R_0$ with a thin hollow core at constant pressure are \cite{barenghi09},
\begin{equation} v_0 = \frac{\Lambda_0\kappa}{4\pi} R_0^{-1},\label{v0}\end{equation}
\begin{equation} E_0 = \frac{(\Lambda_0-1)\rho\kappa^2}{2}R_0 = \frac{(\Lambda_0-1)\rho\kappa^2}{4\pi}L_0,\label{E0}\end{equation}
\begin{equation} I_0 =\pi\rho\kappa R_0^2,\label{I0}\end{equation}
where $\Lambda_0=\ln (8R_0/e^{1/2}a_0) \approx 11$ (for $R_0\sim 1\,\mu$m), $a_0=0.13$\,nm is the vortex core parameter, $\kappa \equiv h/m = 1.00\times 10^{-7}$\,m$^2$s$^{-1}$ is the quantum of circulation ($h$ is Plank's constant and $m$ is mass of a $^4$He atom), $\rho = 145$\,kg\,m$^{-3}$ is the density of liquid $^4$He and $L_0 = 2\pi R_0$ is the ring's circumference. 

The relationships in Eqs.\,\ref{v0}-\ref{E0} were verified in superfluid helium by Rayfield and Reif \cite{rayfield64} in 1964 when quantized vortex rings were first observed. The motion of the rings could be controlled and detected by tagging each ring with a trapped ion. The application of an electric field allowed $R_0$ to be tuned to particular values. These equations have been used to extract values for superfluid $^4$He of $\kappa$ and $a_0$ in the limit of zero temperature and pressure from  measurements of the time of flight (and hence velocity) as a function of the energy imparted to a beam of rings\cite{rayfield64,steingart72}. Subsequently, charged vortex rings have developed into a powerful tool for generating and detecting quantum turbulence \cite{golov09,walmsley14b,zmeev15}.

The paper is organized as follows. In section II we provide a brief overview of the commonly used vortex filament method. The numerical results of the paper are divided between sections III and V; the former describes the interaction between initially unidirectional circular rings, whilst the latter presents the scenario where one of the rings is replaced with a deformed ring to see if the presence of Kelvin wave excitations affects the interaction. In section IV we analyze some properties of single deformed vortex rings.  Finally, in section VI we discuss the implications of these results for experiments in superfluid $^4$He.


\section{Vortex filament method}

Due to the smallness of the core radius $a_0$, on most length scales of interest a quantized vortex can be treated as a line vortex of constant strength $\kappa$ -- an oriented space curve, ${\bf r}={\bf s}(\sigma)$ (where $\sigma$ is the parameter specifying the position along the line, $|d \mathbf{s}| = d \sigma$). 
In the zero-temperature limit, in the absence of dissipative forces, as the mass associated with the vortex core can be neglected, a point $\mathbf{s}$ on a vortex will move with the local superfluid velocity (Helmholtz's theorem) that is given by the Biot-Savart law (for unbounded fluid at rest at infinity), 
\begin{equation}
\mathbf{v}(\mathbf{s})=\frac{\kappa}{4\pi}\oint\frac{\left(\mathbf{s}_1-\mathbf{s}\right)\mathbf{\times}d\mathbf{s_1}}{|\mathbf{s}_1-\mathbf{s}|^3}.
\end{equation}

For the numerical simulations of the dynamics of vortex lines of arbitrary shapes, we use the vortex filament method, following the pioneering work of Schwarz\cite{schwarz85}, to approximate vortex lines as a sequence of connected discrete points: 
\begin{equation}
\mathbf{v}=\frac{\kappa}{4\pi} \mathbf{s}^\prime\times \mathbf{s^{\prime\prime}}\ln\left(\frac{2\sqrt{\xi_+\xi_-}}{e^{1/2}a_0}\right)+\frac{\kappa}{4\pi}\int\frac{\left(\mathbf{s_1}-\mathbf{s}\right)\mathbf{\times}d\mathbf{s_1}}{|\mathbf{s_1}-\mathbf{s}|^3},
\label{vfm}
\end{equation}
where $\xi_+$ and $\xi_-$ are the lengths of the vortex segments connected to point $\mathbf{s}$, $\mathbf{s^\prime}$ is the tangential unit vector and $\mathbf{s^{\prime\prime}}$ points along the principle normal vector with a magnitude $(|\mathbf{s^{\prime\prime}}|)^{-1}$ equal to the local radius of curvature. We have utilized the standard way of desingularizing the Biot-Savart integral which is to extract the local contribution (the first term on the right hand side) leaving the non-local contribution as an integral over all the points that are not connected to $\mathbf{s}$ (the second term).

The local induction approximation (LIA) results from ignoring the contribution of non-local terms and is capable of describing the dynamics of isolated weakly-deformed vortex rings. However, the LIA does not capture the long-range interactions that can occur between vortex rings (such as leapfrogging) nor is it sufficient for accurately describing the effect of large amplitude Kelvin wave excitations on a ring, and thus we use the full Biot-Savart approach, Eq.\,\ref{vfm}, for all of the simulations presented in this paper.

The typical distance between vortex mesh points, $\delta$, is chosen such that each ring consists of $\simeq 100$ points, meaning that $\delta\simeq 80$\,nm for a ring of $1\,\mu$m radius which is typical for the size of rings that occur in superfluid helium experiments. The results presented in the following sections do not change if $\delta$ is decreased and there are always a minimum of six mesh points on each ring. If during a simulation the distance between two points $\xi$ changes such that it is no longer in the range $\delta/2<\xi<2\delta$ then segments are either added (maintaining the local curvature) or removed as required. 

The Biot-Savart formalism is accurate for an ideal incompressible fluid described by the Euler equation, which does not allow vortex reconnections. However, at distances comparable to $a_0$ this description breaks down and vortex lines do reconnect, thus changing the topology of the vortex configuration. Reconnections need to be added by hand to the vortex filament model. This can be achieved in different ways (further details are given in Ref.\,\onlinecite{baggaley12}). We have tried several different reconnection methods, but find that the main details of what we present in this paper do not depend on the precise method used (as was found in simulations of vortex tangles\cite{baggaley12}). In what follows, we reconnect filaments that approach within a distance  $\delta/2$ of each other. The vortex points that are involved in the close approach are removed which ensures that reconnections produce a small loss of line length (in the language of Ref.\,\onlinecite{baggaley12}, these are Type III reconnections).  Further details of how to implement the vortex filament method are described in detail elsewhere\cite{schwarz85,baggaley11,samuels01}.

The following useful properties of a vortex loop can be easily calculated \cite{saffman92}: its length $L$, energy $E$, impulse ${\bf I}$ and angular impulse ${\bf A}$,  
\begin{equation}
L = \oint |d \mathbf{s}|,
\label{L}
\end{equation}
\begin{equation}
E=\rho\kappa\oint\mathbf{v}\cdot\mathbf{s}\times d\mathbf{s},
\label{Ef}
\end{equation}
\begin{equation}
\mathbf{I}=\frac{1}{2}\rho\kappa\oint\mathbf{s}\times d \mathbf{s},
\label{I}
\end{equation}
\begin{equation}
\mathbf{{\bf A}}=-\frac{1}{2}\rho\kappa\oint s^2 d \mathbf{s}.
\label{A}
\end{equation}
It is convenient to define the effective radius $R'$ as the radius of a circular vortex ring that has the same impulse as the ring under question, using Eqs.\,\ref{I0}, 
\begin{equation}
R' = \left(\frac{I}{\pi \rho \kappa}\right)^{1/2},
\label{Reff}
\end{equation}
The energy $E$ of weakly-deformed vortex ring, according to the LIA, is approximately proportional to its length $L$, 
\begin{equation}
E_L = \frac{(\Lambda-1)\rho\kappa^2}{4\pi}L,
\label{E}
\end{equation}
where $\Lambda \approx \Lambda_0(R_0)$ as in Eq.\,\ref{E0} with $R_0\approx R'$.
For large-amplitude deformations, Eq.\,\ref{E} gives an approximate value for the total energy (Eq.\,\ref{Ef}) because it ignores the additional energy of flow induced by the non-local term in Eq.\,\ref{vfm}. For the simulations in this paper, Eq.\,\ref{E} gives an accurate estimate of the total energy because the deformations consist of a broad distribution of Kelvin wave modes such that the total non-local contribution to the total energy is small (as both constructive and destructive interference due to non-local effects are minimal). 

A weakly-deformed vortex ring can thus be characterized by its (i) energy, i.\,e. length $L$; (ii) impulse (i.\,e. the effective radius $R'$ and direction of propagation  and (iii) position of the geometrical centre ${\bf r}_0(t)$ (and hence, velocity $d {\bf r}_0/ d t$). The difference between the total energy $E$ and the energy $E_0(R')$ of an effective smooth ring of radius $R'$ gives the energy and effective amplitude of deformations (Kelvin waves). When deformations are small, the smooth radius of the ring $R_0$, position $\mathbf{r}_0$ and spectrum of Kelvin waves can also be calculated approximately by projecting the vortex line $\mathbf{s}(\phi)$ (where $\phi$ is the azimuthal angle in the plane perpendicular to the direction of propagation), via a Fourier transform, on a circular ring with superimposed harmonic helical waves of all allowed wavenumbers \cite{RistoFFT13}. As we discuss in Section IV, within the small range of fluctuations of $R_0$ due to standing Kelvin waves, $R' \approx R_0$. 
  
 Strong deformations can lead to coiled structures\cite{baggaley11} that contribute a component of impulse in the opposite direction \cite{ricca13} and would cause our extracted value of $R'$ to underestimate the smoothed radius of the underlying ring. For this reason, we only use values of $R'$ calculated for rings in their final stable state with sufficiently small deformation (as the large-amplitude deformations that can be created when two rings reconnect tend to quickly produce a self-reconnection and the emission of a small ring leaving behind rings with deformations that are mainly helical in nature). 
This also means $R'$ is extracted for rings that have moved far away from each other (typically $> 10 R'$) where non-local effects from other vortices are negligible and they can thus be considered as stable independent entities. 

\section{Interactions between circular rings}
We have simulated the interaction of two circular rings with initial radii $R_1$ and $R_2$ that are traveling initially in the $z$-direction but with their axes offset by an impact parameter, $b$. The larger ring ($R_1\geq R_2$) is always placed a distance $d$ in front of the smaller ring such that the initial coordinates $(x,y,z)$ for the centers of the rings at time $t=0$ are (0,0,$d$) and ($b$,0,0). The mean initial radii of the rings was kept fixed, $R_{\mathrm{m}}=\left(R_1+R_2\right)/2=0.95\,\mu$m, but the difference in radii, $\Delta R=R_1-R_2$, was the second parameter that was varied. 
The separation $d=5\,\mu$m ($\gg R_{\mathrm{m}}$) was used so that the rings can, as a good approximation, be considered to be independent non-interacting rings at $t=0$. Using larger values of $d$ up to 10\,$\mu$m barely made any difference to the outcomes described below. The ring configurations were allowed to evolve for a total time of $1-2$\,ms, by which point the rings had interacted and moved far away from each other, and values of $R'$ could be reliably extracted.

\begin{figure}
\includegraphics[width=8cm]{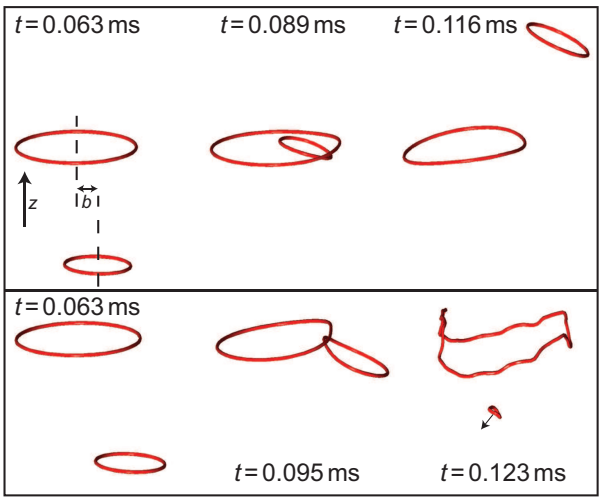}
\caption{A sequence of snapshots of the vortex filaments for two different examples of vortex ring interactions with time increasing from left to right. In both cases, the rings are initially moving vertically upwards (in the $z$-direction) with initial radii  $R_1=1.2\,\mu$m and $R_2=0.7\,\mu$m ($R_{\mathrm{m}}=0.95\,\mu$m, $\Delta R=0.5\,\mu$m) and a vertical separation of $d=5\,\mu$m. Top panel:  impact parameter, $b=0.4\,\mu$m. The smaller ring passes through the center of the larger ring without reconnecting. Bottom panel: impact parameter, $b=0.95\,\mu$m.  This time the smaller ring clips the outside of the larger ring resulting in a reconnection and the formation of a single ring. The reconnection produces large deformations of the ring in the form of Kelvin waves. A subsequent self-reconnection leads to the emission of a very small ring. All rings continue to have a positive $z$-component of velocity unless indicated otherwise with an arrow (the small ring in the bottom right snapshot). For each snapshot, the position in the $z-$direction is shifted to maintain the larger ring in the same place. \label{fig_snapshots}}
\end{figure}

Two examples of interacting rings are shown in Fig.\,\ref{fig_snapshots} at three different times. The smaller (and thus faster) ring begins to catch up with the larger ring, and the long-range flow fields of the rings cause them to begin interacting. This non-local effect causes the vortex ring in front to increase in size while the ring at the back gets smaller. They also tend to repel each other sideways. The top panel of Fig.\,\ref{fig_snapshots} shows an example of a small but finite value of the impact parameter when the smaller ring passes through the other ring without reconnecting. Both rings acquire a transverse component of velocity and they subsequently move apart with their axes no longer aligned. In contrast, the bottom panel is an example with a larger $b$ that leads to the rings colliding and reconnecting at $t=0.097$\,ms, resulting in the formation of a large ring, which in this case, at $t=0.119$\,ms, emits a small ring due to a self-reconnection. The large ring is clearly non-circular, with the distorted shape arising due to the propagation of large-amplitude Kelvin waves around the ring from the sharp cusps created by the initial reconnection \cite{hanninen13,hanninen15}.

\begin{figure}
\includegraphics[width=8cm]{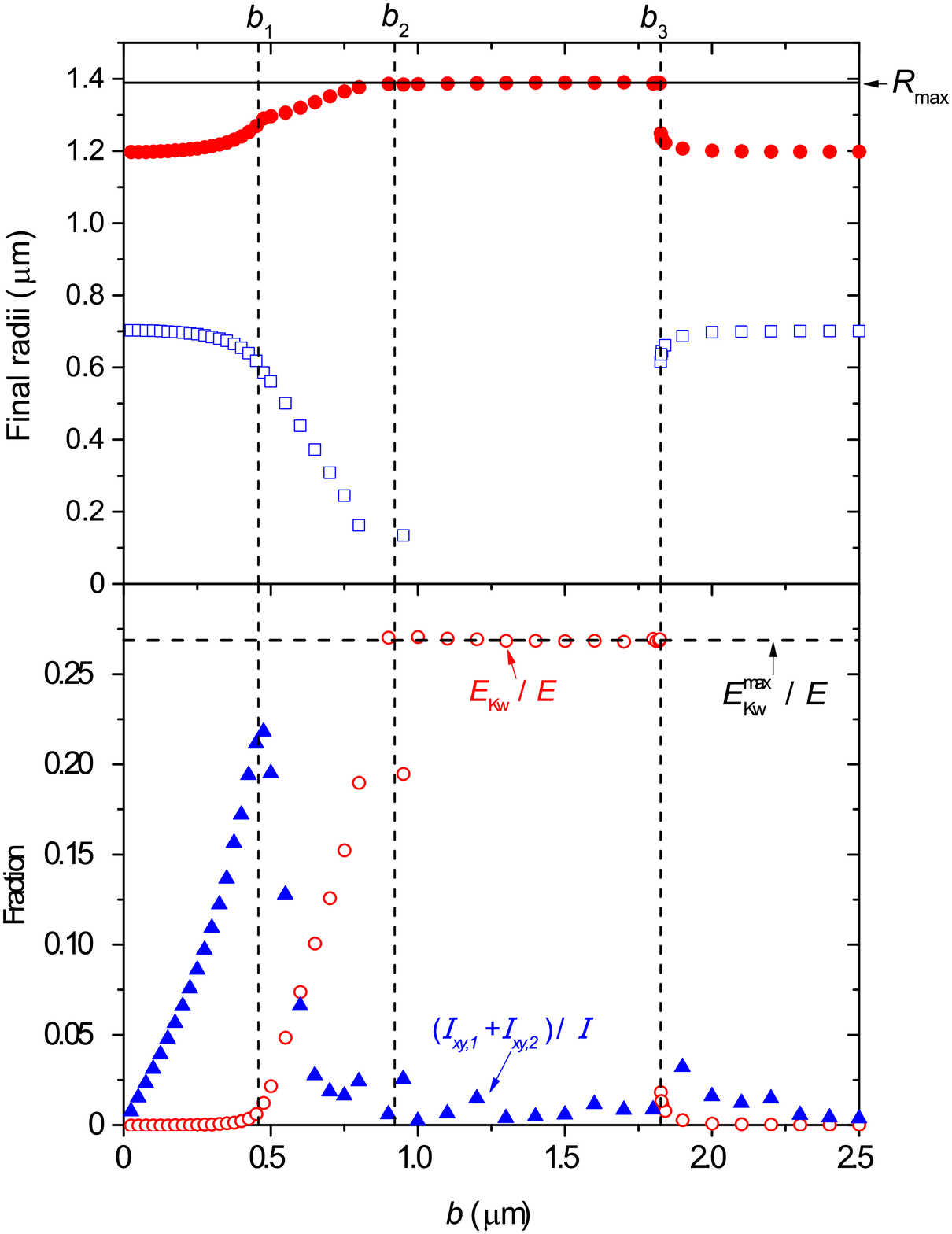}
\caption{Top panel: The effective final post-interaction radii of the rings, $R'$ versus impact parameter, $b$. The initial radii of the rings were $R_1=1.2\mu$m and $R_2=0.7\mu$m with $d=5\mu$m. The vertical dashed lines denote the transitional impact parameters at the boundaries between the different regimes described in the main text. The solid horizontal line indicates the maximum possible radius, $R_\mathrm{max}=\sqrt{R_1^2+R_2^2}$, obtained from the conservation of impulse (Eq.\,\ref{ConservedI}) when two rings merge into a single ring. Bottom panel: The fraction of energy converted into deformations (open red circles) and the sum of the transverse components of impulse normalized by the total impulse (closed blue triangles) versus impact parameter. The dashed horizontal line indicates the maximum possible fraction of energy due to deformations, $E_{\rm Kw}^{\rm max}/E = 1-R_\mathrm{max}/(R_1+R_2)$, when two rings merge into a single ring. \label{fig_radii}}
\end{figure}

In what follows, we will denote the final effective radii of isolated post-interaction (daughter) rings as $R_  i'$. 
The values of $R'$ for $\Delta R=0.5\,\mu$m are shown in the top panel of Fig.\,\ref{fig_radii} as a function of $b$. We note that $R'(b)$ for the smaller ring is qualitatively similar to that observed in recent simulations of the interaction between a vortex ring and an initially straight vortex  line \cite{laurie15,villois15}. Further insight is provided by the amount of energy converted from translational motion of the ring into deformations. As the total energy $E$ is conserved (losses due to reconnections and numerical dissipation at length scale $\sim \delta$ are negligible, see Section IV), and if the translational energy of the rings in their final state is obtained by inserting $R'$ into Eq.\,\ref{E0}, then the fraction of energy associated with deformations is
\begin{equation}
\frac{E_\mathrm{Kw}}{E} \approx 1-\frac{\sum_i R_i'}{R_1+R_2},
\end{equation}
where the summation is over all rings in the final state. 
As the total impulse is conserved, we can relate the effective radii of initial and daughter ($R_i'$) vortex rings:
\begin{equation}
R_1^2+R_2^2 = \sum_i R_i'^2 .
\label{ConservedI}
\end{equation}
As well as creating deformations, the interaction between the rings can also produce transverse motion of the rings (even without reconnections). This is quantified from the sum of the magnitudes of the transverse components of impulse for each ring in their final state, $I_{xy}=\sqrt{I_x^2+I_y^2}$. The dependence on $b$ of both $I_{xy}$ and $E_\mathrm{Kw}$ is shown in the bottom panel of Fig.\,\ref{fig_radii}.

We have identified four different regimes, each of which is marked on Fig.\,\ref{fig_radii}, that depend on the impact parameter as follows:
\begin{enumerate}
\item $b<b_1$, the second ring passes through the first ring without reconnecting (as shown in the top panel of Fig.\,\ref{fig_snapshots}). There is no significant energy associated with deformations of the rings (although there is some slight quadrupolar distortion of both rings). The values of $R'$ are comparable to the initial radii. The rings tend to fling each other sideways with the amount of transverse motion increasing as $b$ increases.

\item $b_1< b<b_2$, the rings reconnent in two points, producing two rings that begin to have noticeably dissimilar $R'$ as $b$ increases, with $R_2'\rightarrow 0$ as $b\rightarrow b_2$. The fraction of energy converted to deformations begins to increase although the transverse components of impulse now decrease (there is thus a maximum at $b_1$).

\item $b_2< b< b_3$, the rings reconnect once and thus merge together producing a large single deformed ring which can sometimes emit a small ring at a later time due to a self-reconnection  (as shown in the bottom panel Fig.\,\ref{fig_snapshots}). The effective radius of the large ring confirms that impulse is conserved in the merger:  the radius of the ring is equal to $R_\mathrm{max}=\sqrt{R_1^2+R_2^2} = \sqrt{2}R_\mathrm{m}\sqrt{1+(\Delta R/2R_\mathrm{m})^2}$ (using Eq.\,\ref{ConservedI}), which tends to $\approx \sqrt{2}R_{\mathrm{m}}$ for $\Delta R \rightarrow 0$. The fraction of energy converted into deformations is constant, $E_\mathrm{Kw}^\mathrm{max}/E=1-R_\mathrm{max}/2R_\mathrm{m}$. Its largest value is in the limit $\Delta R  \rightarrow 0$,
\begin{equation}
E_\mathrm{Kw}^\mathrm{max}/E = 1-2^{-1/2} = 0.29.
\label{X}
\end{equation} 

\item $b> b_3$, the rings miss each other on the outside with very little change in the shape, radii and direction of the rings.
\end{enumerate}

The range of impact parameters where the rings reconnect is $b_1< b < b_3$. The transitions at each of the bounding limits are noticeably different. There is a continuous variation of $R'$ for both rings at $b_1$ whereas the transition at $b_3$ is nearly discontinuous. The case of $b\simeq b_1$ is essentially the reconnection of two initially parallel vortex lines whereas $b_2 \leq b \leq b_3$ corresponds to the case of vortex lines that are initially antiparallel. We note that in other simulations the reconnection of antiparallel lines produces a cascade of small vortex rings\cite{svistunov95,kerr11,kursa11}.  

\begin{figure}
\includegraphics[width=8cm]{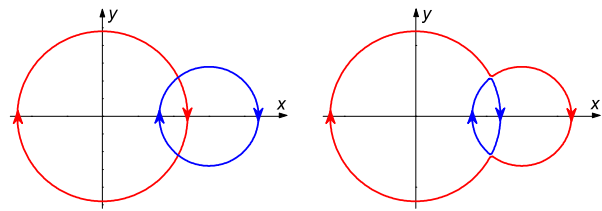}
\caption{(left) $(x,y)$ coordinates of vortex rings, based on the LIA, just before reconnection. (right) New topology ater reconnection. $\Delta R = 0.5$\,$\mu$m, $R_{\rm m} = 0.95$\,$\mu$m, $b = 1.5$\,$\mu$m.} 
\label{LIA_cartoon}
\end{figure}

\begin{figure}
\includegraphics[width=8cm]{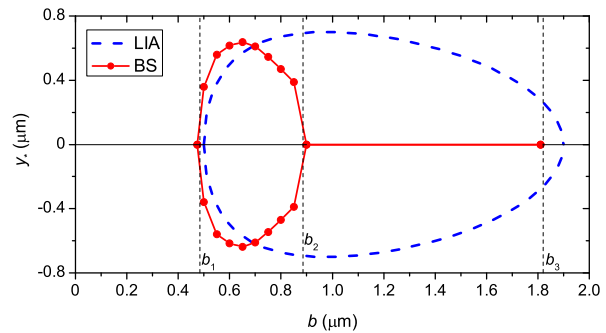}
\caption{Locus of $y$-coordinates at reconnection for various impact parameters $b$ for LIA (dashed blue line) and full Biot-Savart (solid red line) methods. $\Delta R = 0.5$\,$\mu$m, $R_{\rm m} = 0.95$\,$\mu$m. Values of $b_1$, $b_2$ and $b_3$, separating different regimes of reconnections, are shown by vertical dashed lines.} 
\label{ReconnectionPointVsImpactPar}
\end{figure}

\begin{figure}
\includegraphics[width=8cm]{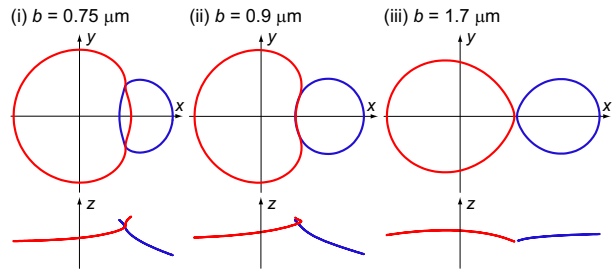}
\caption{The $xy$ (top) and $xz$ (bottom) projections of vortex rings, calculated using the full Biot-Savart method (Eq.\,\ref{vfm}), just before reconnection: (i) $b=0.75\,\mu$m $(b_1<b<b_2)$, (ii) $b=0.9\,\mu$m $(b\approx b_2)$ and (iii) $b=1.7\,\mu$m $(b\approx b_3)$. $\Delta R=0.5\,\mu$m, $R_{\rm m}=0.95\,\mu$m. There are two reconnections in (i) that create two daughter rings but only one reconnection at $y=0$ in (ii) and (iii) leading to the formation of a single deformed ring.}
\label{fig_BS_snapshots}
\end{figure}

The very existence of the third regime (in which only one reconnection occurs for a range of impact parameters $b_2 < b < b_3$), is a property of the Biot-Savart solution; no such range of $b$ exists for the LIA (for which $b_2=b_3$). Our initial conditions are mirror-symmetric with respect to the $x-z$ plane, hence, all subsequent evolution maintains this symmetry (any small deviations stem from the numerical implementation).  In the LIA model, circular rings travel, without changing shape, all the way until instantaneously reconnecting at two points $(x_*, y_*, z_*)$ and $(x_*, -y_*, z_*)$ (see example in Fig.\,\ref{LIA_cartoon}). There are generally two reconnections, except for the singular cases, $b=R_1-R_2$ and $b=R_1+R_2$, for which both reconnections converge into a single point $y_*=0$ at $x_*=R_1$. 
The full locus $(x_*,y_*)$ vs. $b$ for LIA $\Delta R = 0.5$\,$\mu$m and $R_{\rm  m} = 0.95$\,$\mu$m is shown in Fig.\,\ref{ReconnectionPointVsImpactPar} by dashed line. 
However, for our Biot-Savart calculations, shown by solid line in Fig.\,\ref{ReconnectionPointVsImpactPar}, the right point with $y_*=0$ stretches into a range of $b$ ($b_2 < b < b_3$), whereas the left ($b_1$) remains a singular point. Snapshots of ring configurations just before reconnections are shown in Fig.\,\ref{fig_BS_snapshots} for three values of $b$: one just below $b_2$ and two others between $b_2$ and $b_3$. In each case, regions of close approach of rings develop cusps which stretch towards each other and then reconnect. 
It seems the origin of this effect is similar to that of Crow instability \cite{Crow} of two antiparallel vortex lines with respect to reconnecting. In our case of initially circular  antiparallel vortex lines, a small initial overlap (larger $b$) of rings seems to result in a single point of close approach at $y_*=0$, while greater overlaps (smaller $b$) eventually favor two distinct points of close approach at $\pm y_* \neq 0$, from which the cusps begin to grow.

\begin{figure}
\includegraphics[width=8cm]{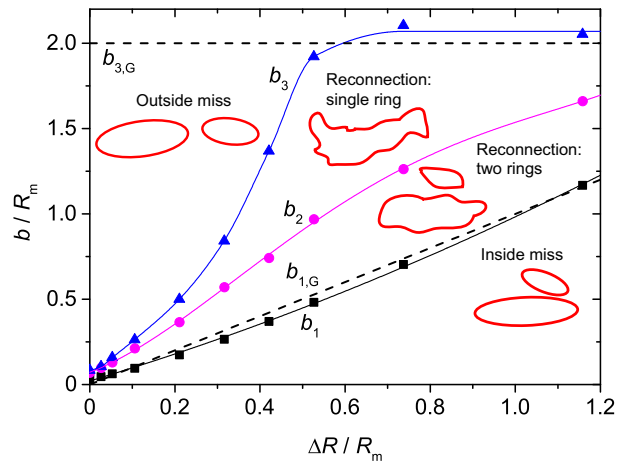}
\caption{Transitional impact parameters versus the difference in initial radii $\Delta R$ with both axes normalized by the mean initial radii $R_{\mathrm{m}}$. The dashed lines indicate the values $b_{1,G}$ and $b_{3,G}$ based on geometric expectations of where a reconnection should occur (i.e. if the LIA were used). The solid lines are guides to the eye. The cartoons illustrate the type of interaction that occurs in each region. \label{fig_phase}}
\end{figure}

The outcomes described above, both in terms of final radii and re-distribution of energy, are features that are common to all interactions between unidirectional circular rings irrespective of $\Delta R$. However, the transitional impact parameters ($b_1,b_2$ and $b_3$) do vary with $\Delta R/R_{\mathrm{m}}$ as shown in Fig.\,\ref{fig_phase} which is essentially the ``phase diagram". All three transitional values of $b$ increase as $\Delta R$ increases. However, when the rings have nearly the same initial radius  (and hence a small relative velocity and long interaction time), the first three regimes are very narrow, and only a small value of $b_3/R_{\mathrm{m}}=0.084$ is required before the rings are able to repel each other sufficiently sideways so that there is no reconnection. This is in contrast to the interaction of strongly dissimilar rings where there is thus a much broader range of impact parameters resulting in a reconnection. The upper and lower values of the impact parameter, $b_{1,\mathrm{g}}$ and $b_{3,\mathrm{g}}$, where a reconnection would be expected from a geometrical extrapolation of the trajectories of initial circular rings (which would be the solutions of the LIA), are also shown in Fig.\,\ref{fig_phase} by dashed lines. It seems, the observed values  $b_1$ are close to the geometrical ones, $b_{1,\mathrm{g}}$, at any $\Delta R/R_\mathrm{m}$. On the other hand, $b_3(\Delta R/R_\mathrm{m})$ has two clear limits, roughly separated at $\Delta R/R_\mathrm{m} = 0.5$. At larger $\Delta R/R_\mathrm{m}>0.5$, the borderline values of impact parameters $b$ are close to those expected from geometrical cross sections, while at smaller $\Delta R/R_\mathrm{m}<0.5$, these borderline values of $b_3$ tend to a very small value (perhaps, zero) in the limit $\Delta R \rightarrow 0$ -- when the two rings have sufficient time to repel each other sideways and thus avoid any reconnection.

The tendency to create vortex rings of highly unequal size upon a collision of two unidirectional rings can be attributed to constraints arising from the conservation laws. 
First let us consider a head-on collision of two circular rings of equal radii $R$, which results in two daughter vortex rings of effective radii $R_1'$ and $R_2'$. As the total energy (Eq.\,\ref{E}) and impulse (Eq.\,\ref{Reff}) are conserved, assuming $\Lambda = {\rm const}$, we have
\begin{equation}
2R > R_1' + R_2'
\label{EnergyHeadOn}
\end{equation}
 and 
 \begin{equation}
0 = R_1'^2-R_2'^2.
\label{ImpulseHeadOn}
\end{equation} 
The former inequality arises from the fact that some part of energy becomes the energy of deformations of the daughter rings (and another small part is lost during the reconnection), while the latter equality indicates that the center of mass remains at rest. Hence, secondary rings of equal (albeit slightly smaller due to Eq.\,\ref{EnergyHeadOn}, $R_1'=R_2' < R$) radii have to be created -- propagating in opposite directions. 

For another limit, of the collision of two unidirectional circular vortex rings of similar radii $R_1 \approx R_2 \approx R$, that results in two daughter deformed vortex rings of effective radii $R_1'$ and $R_2'$, we have
\begin{equation}
2R > R_1' + R_2'
\label{EnergyUnidirec}
\end{equation}
 and 
 \begin{equation}
2R^2 \leq R_1'^2+R_2'^2.
\label{ImpulseUnidirec}
\end{equation} 
The latter becomes an equality if the daughter rings propagate in the same direction as the primary ones. These equations do not have solutions $R_1'=R_2'$, thus only allowing daughter rings of unequal size, e.\,g. $R_1' > R$ and $R_2' <R$. The more energy that is transferred to Kelvin waves in Eq.\,\ref{EnergyUnidirec}, the more unequal are the radii of the secondary rings. 
Another possible outcome is the formation of just one large secondary ring, i.\,e. $R_1' = \sqrt{2}R$, $R_2'=0$. This has the largest possible deformations that might cause self-reconnections and the subsequent shedding-off of one or more tiny secondary rings of radius $R' \ll R$ in arbitrary directions (which would not strongly affect the balance of total impulse due to their small size). 

We thus conclude that quick head-on collisions favor the production of daughter vortex rings of similar size and opposite direction of propagation (the vortex rings ``pass through each other'').  In the opposite case of unidirectional vortex rings, their collisions must become less likely due to the long time of approach when $\Delta R \ll R$. When unidirectional rings do reconnect, strongly unequal daughter rings are favored; this includes the limiting case of the rings merging to form one large deformed ring carrying the total impulse. Our numerical simulations of interacting unidirectional rings revealed examples of all of these features. 

\begin{figure}
\includegraphics[width=8cm]{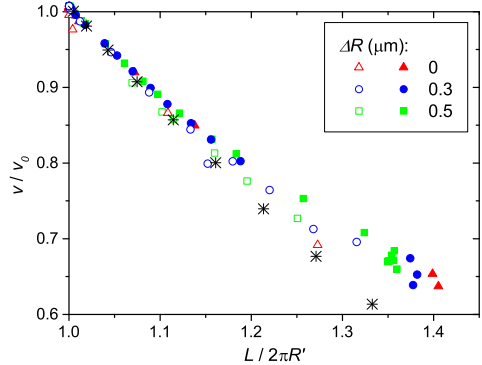}
\caption{Ratio of the velocity of vortex rings post-interaction to $v_0(R')$ from Eq.\,(\ref{v0}) versus level of non-circular distortion for all values of $b$ and three different values of $\Delta R$. The open (closed) symbols are for the smaller (larger) rings created following the interaction. Asterisks show the results of simulations of the dynamics of a circular ring with one helical wave of wavenumber $N=4$ and variable amplitude.
\label{velocity}}
\end{figure}

After interacting, the velocity of deformed rings $v$ can be substantially reduced below that given by inserting $R'$ into Eq.\,\ref{v0}, $v_0$. The effect of Kelvin waves reducing the velocity of vortex rings has been studied extensively \cite{kiknadze02,barenghi06,helm11,sonin12}. In Fig.\,\ref{velocity}, the velocities of vortex rings for several different initial $\Delta R$ are plotted against the total line length of each ring, $L$, scaled by the circumference of the equivalent circular ring, $2\pi R'$. The main observation is the existence of a universal curve that does not depend on $b$ or $\Delta R$. We also confirm that the velocity reduction is the greatest for the most highly distorted rings (created when $b_2\leq b\leq b_3$).

Previous numerical and analytical work \cite{kiknadze02,barenghi06,helm11,sonin12} focused on the change in velocity of vortex rings when there is regular periodic arrangement of Kelvin waves of large amplitude. Our own numerical simulations of vortex rings with a single harmonic deformation of amplitude $A$ and wavelength $2\pi R_0/N$ (with integer $N\geq 2$) resulted in velocities equal to those calculated by Barenghi {\it et al.} \cite{barenghi06,helm11}; some example data for $N=4$ are shown in Fig.\,\ref{velocity}. There is agreement with the velocities for arbitrarily deformed vortex rings for $1\leq L/2\pi R' \lesssim 1.2$, while for $L/2\pi R'\gtrsim 1.2$ arbitrarily deformed rings show a smaller effect on $v/v_0$ than those with a single Kelvin wave mode. As the excitation of a single Kelvin-wave mode may prove difficult to obtain experimentally, we suggest that the curve in Fig.\,\ref{velocity} could be used as the signature of anomalously slow vortex rings that can be relatively easily created by merging rings together.

\begin{figure}
\includegraphics[width=8cm]{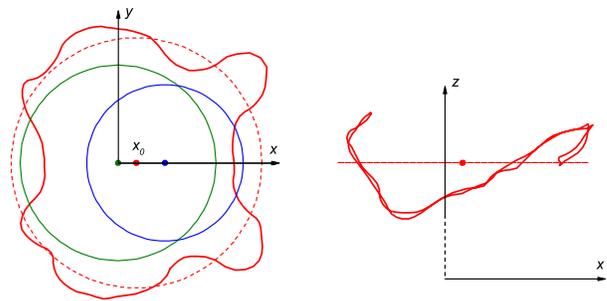}
\caption{
The $xy$ and $xz$ projections of the distorted ring created by colliding two circular rings, both initially in $xy$-plane: one centered at (0, 0, 3\,$\mu$m) with $R_1=1.0\,\mu$m (green) and the other at (0.48\,$\mu$m, 0, 0) with $R_2=0.8\,\mu$m (blue). The dashed lines correspond to a ring with the smooth effective radius $R'=1.28\,\mu$m, centred at $(x_0, 0, z_0)$ with $x_0 = 0.19\,\mu$m and $z_0 = 11.73\,\mu$m. The scales in all directions $(x, y, z)$ are the same. The approximate symmetry with respect to reversing the $y$- component is evident: if point $(x,y,z)$ belongs to the deformed ring, the point $(x,-y,z)$ also belongs to the ring. 
}
\label{disring}
\end{figure}

\begin{figure}
\includegraphics[width=8cm]{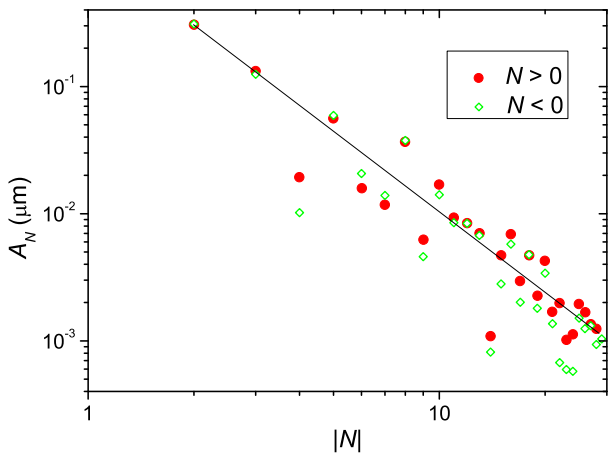}
\caption{Spectrum of the Fourier amplitudes $A_N$ of the vortex line ${\bf s}(\phi)$ (where $\phi$ is the azimuthal angle in $xy$ plane with origin $(x_0,0)$), resulted from a  reconnection of circular rings with $R_1 = 1.2\,\mu$m, $R_2=0.8\,\mu$m and $b=1.5\,\mu$m. Time $t=0.741$\,ms. The dashed line $A_N \propto N^{-2.1}$ guides the eye.}
\label{FFT}
\end{figure}

\section{Properties of deformed vortex rings}

The most strongly deformed but stable rings in our simulations are those obtained after a merger of two rings into one large ring, when $b_2 < b < b_3$. In this Section we consider this interesting case. 

The conservation of impulse ensures that the daughter ring travels in the $z-$direction with $R'=R_\mathrm{max}$ and the fraction of energy associated with Kelvin wave deformations, $E_{\mathrm{Kw}}^{\mathrm{max}}/E\approx 0.29$ (Eq.\,\ref{X}). According to Eq.\,\ref{E}, $E/E_0 \approx L/(2\pi R_0)$, so this value can be compared with the relative increase in the length of the most strongly deformed rings in our simulations (as in Fig.\,\ref{velocity} for $\Delta R \rightarrow 0$), $E_{\mathrm{Kw}}/E=1-2\pi R'/L=0.29$. We thus find these two estimates of the excess energy and line length are in good agreement; the fact that both are substantially smaller than unity gives footing to the use of the small-amplitude approximation for our analysis.

The conserved angular impulse gives the position $(x_0,0)$, in the $xy$ plane, of the centre of the daughter ring:
\begin{equation}
x_0 = \frac{R_2^2}{R_1^2+R_2^2}b.
\label{ConservedA}
\end{equation}

During the typical run of length $\simeq 1$\,ms, the calculated impulse was constant, as expected. The mean total length $L$ was also constant within our accuracy and equal to $0.99 \times 2\pi (R_1 + R_2)$. This gives the lower bound on the total energy $E$, which is hence not more than only 1\% smaller than the total energy of the two initial rings. The same result was also obtained using Eq.\,\ref{Ef}. We thus conclude that, in our model, the energy loss due to a single reconnection does not exceed 1\%, and the subsequent dissipation (apparently due to the direct energy cascade through the non-linear interactions of Kelvin waves without reconnections), if any, is much smaller.

An example of the shape of such a deformed ring is shown in Fig.\,\ref{disring}. Because of the mirror symmetry with respect to changing the sign of the $y$-coordinate, the resulting deformed ring undergoes oscillations (see Fig.\,\ref{disring}, in which the projections of the vortex line for its two half-rings coincide); in other words, the deformations  are standing waves composed of equally-populated Kelvin-wave modes of opposite directions of propagation. A Fourier transform reveals that for all $N$ such that $|N| \lesssim 20$, the corresponding components have nearly equal amplitudes, $A_{|N|} \approx A_{-|N|}$ and are almost perfectly in-phase. In Fig.\,\ref{FFT}, the Fourier transform \cite{RistoFFT13} of the shape of the ring, generated by the reconnection of two rings with $R_1 = 1.2\,\mu$m, $R_2=0.7\,\mu$m and $b=1.5\,\mu$m, shows the spectrum that generally follow $|A_N| \propto N^{-\epsilon}$ with $\epsilon \approx 2$ which is dominated by the standing wave of the fundamental mode $|N|=2$ of amplitude $2A_2 = 0.61\,\mu$m, and hence the relative amplitude of $2A_2/R_0=0.42$. 
The zero-order ($N=0$) Fourier component yields the radius of the ``backbone'' circular ring, $R_0$, that oscillates between circa $1.38\,\mu$m and $1.52\,\mu$m which overlap with  $R' = 1.39\,\mu$m. The value of $\epsilon$ is slightly larger than the exponents expected in various models of the non-linear cascade of Kelvin waves (with $\epsilon=17/10=1.7$ (Ref.\,\onlinecite{kozik04}), $\epsilon=11/6\approx 1.83$ (Ref.\,\onlinecite{lvov10}), and $\epsilon=3/2=1.5$ (Refs.\,\onlinecite{VinenTsubota2003,SoninPRB2012}), although in our case the spectrum appears without the need for a cascade \cite{hanninen15}.

With a single helical wave of wavenumber $N$ of small amplitude $A$ ($A \ll R_0$ and $A^2N^2 \ll 2R_0^2$), superimposed on a circular ring of radius $R_0$, the total length $L$ of the vortex line is
\begin{equation}
\frac{L}{2\pi R_0} \approx 1 + \frac{A^2}{R_0^2}\left(\frac{1}{4} + \frac{N^2}{2}\right).
\label{LDeformed}\end{equation}
As for the most strongly deformed rings $L/(2\pi R') = 1.4$ (Fig.\,\ref{velocity}), using the dominant $N=2$, one arrives at the estimate $A/R_0 = (0.4/2.25)^{1/2} = 0.42$, in good agreement with the result of the Fourier-series analysis.

\section{Interaction with a deformed ring}

In section III we focused on the collision of two initially circular rings. However, experiments can often consist of a cloud of many vortex rings. Thus, when a large slowly-moving deformed ring is created through the merger of two smaller rings, it is likely it will then get hit from behind by faster-moving small rings. It may be that these secondary interactions, when at least one of the colliding rings is already strongly deformed, are more likely to produce small rings. We have simulated this process by firstly merging two circular rings ($R_1=1\,\mu$m, $R_2=0.8\,\mu$m, $d=3\,\mu$m and $b=0.48\,\mu$m) to create a deformed ring of effective radius $R'=R_\mathrm{max} = (R_1^2+R_2^2)^{1/2} = 1.28\,\mu$m (Eq.\,\ref{ConservedI}) with its center positioned at $x_0 = \frac{R_2^2}{R_1^2 + R_2^2}b = 0.19\,\mu$m (Eq.\,\ref{ConservedA}), shown in Fig.\,\ref{disring}. We checked that this deformed ring was stable by allowing it to evolve on its own for a further 10\,ms; during this time it traveled a distance of $350\,R'$ without any self-reconnections taking place. The mean radius of curvature of the ring was 0.4\,$\mu$m. 

We then place a circular ring, of radius 0.8\,$\mu$m, a distance of 2\,$\mu$m behind the deformed ring. The impact parameter, relative to the center of mass of the larger deformed ring, was varied in two dimensions, $\mathbf{b}=(b_x,b_y)$, due to the lack of axial symmetry.

\begin{figure}
\includegraphics[width=8cm]{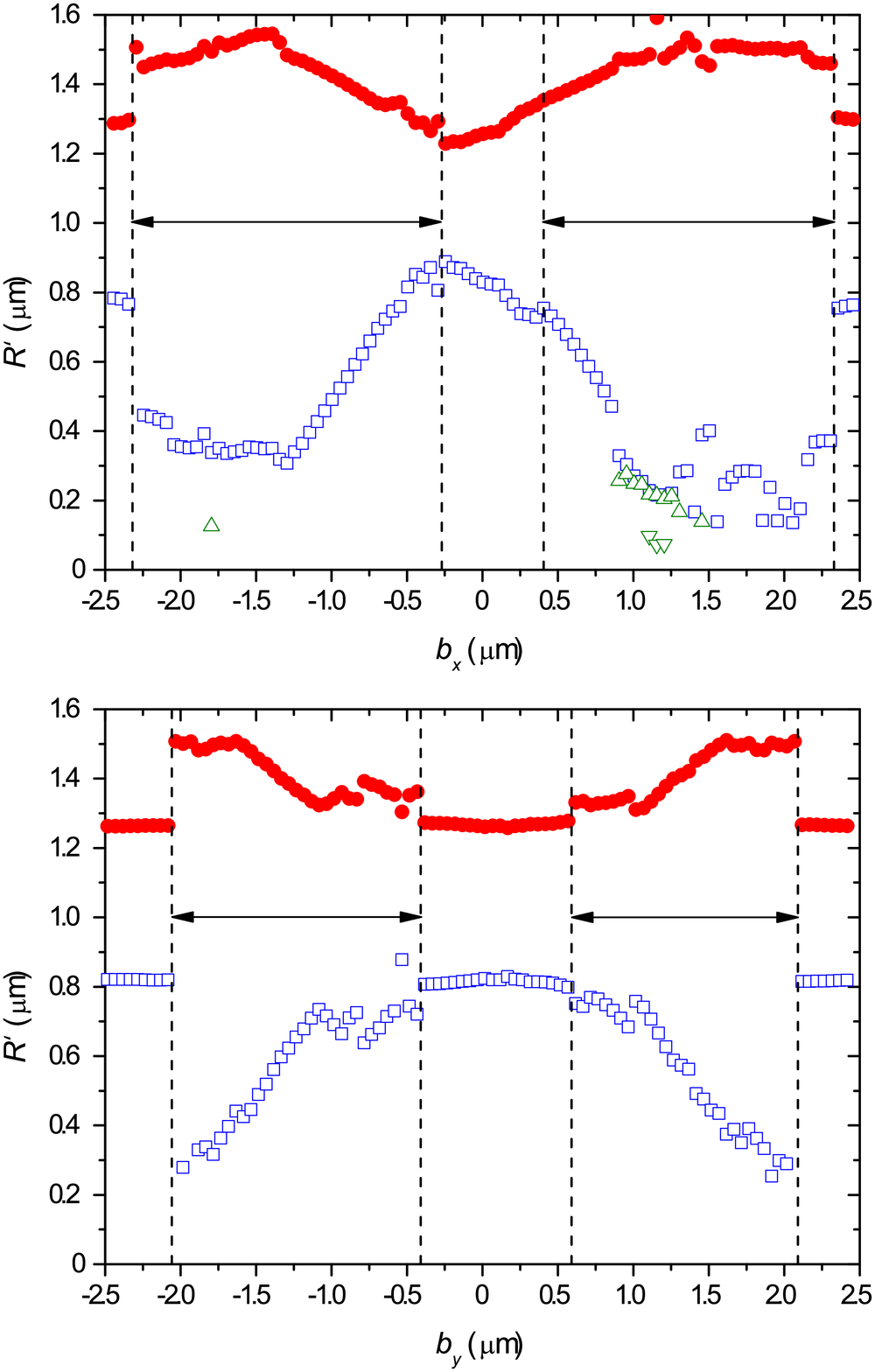}
\caption{The effective post-interaction radii versus impact parameter for $\mathbf{b}=(b_x,0)$ (top panel) and $\mathbf{b}=(0,b_y)$ (bottom panel)  following an interaction of a circular ring of radius $0.8\,\mu$m that is initially placed $2\,\mu$m behind the large ring shown in Fig.\,\ref{disring}. The green open triangles indicate the production of additional small rings (up and down triangles indicate the second and third rings respectively). The arrows and dashed lines indicate the range of impact parameters where reconnections occur. \label{fig_disringxy}}
\end{figure}

The values of $R'$ when $b$ is varied along the $x$ and $y$ axes are shown in Fig.\,\ref{fig_disringxy} (top) and (bottom), respectively. There are clear differences between the two plots because the transitional impact parameters are now functions of both $x$ and $y$ although many of the general features presented in the previous section are still present. The most notable difference is that the formation of a single merged ring (regime 3) is now highly unlikely with a small ring being a more likely result. Indeed, for $b_x\simeq 1\mu$m and $b_y=0$ (Fig.\,\ref{fig_disringxy} (top)) it is possible for two or three small rings to be emitted following the collision.

\begin{figure}
\includegraphics[width=8cm]{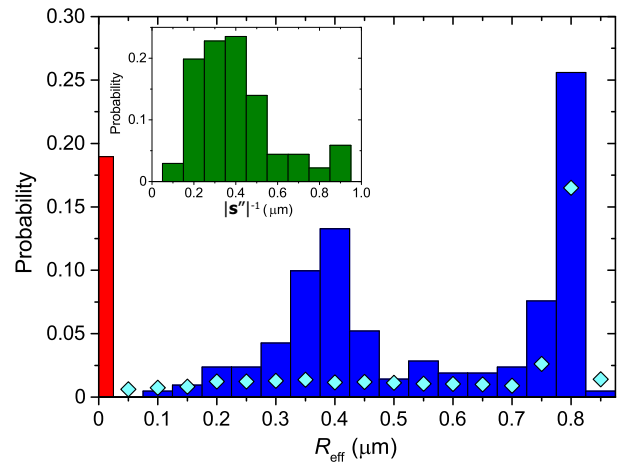}
\caption{Probability distribution for the effective radii of small rings after a circular ring (with initial radius $0.8\,\mu$m) has interacted with the deformed ring shown in Fig.\,\ref{disring}. The impact parameter was randomly selected in the range $(-2.5\,\mu\mathrm{m},-2.5\,\mu\mathrm{m})\leq (b_x,b_y)\leq (2.5\,\mu\mathrm{m},2.5\,\mu\mathrm{m})$ and a total of 211 realizations of the simulation were run. The red bar on the left indicates the probability that a large single ring is formed without the creation of a small ring. The diamonds indicate the probability of the same process but for the interaction with a circular vortex ring of radius $R'$ with otherwise identical parameters. In this case, the probability of a large single ring forming is much larger at 0.64. Inset: the distribution of the initial radii of curvature $(|\mathbf{s^{\prime\prime}}|)^{-1}$ for the mesh points on the deformed loop. \label{fig_prob}}
\end{figure}

In order to probe the probability of creating small rings, the impact parameter was varied with both $b_x$ and $b_y$ chosen randomly in the range $\pm 2.5\,\mu$m for 211 separate runs of the simulation with otherwise identical parameters. The probability distribution of $R'$ of the smaller rings is shown in Fig.\,\ref{fig_prob}. The large peak with $R'\simeq 0.8\,\mu$m are rings that have not reconnected with the deformed ring. There is broad distribution of rings with smaller radii, produced by reconnections, with a peak at $\simeq 0.4\mu$m. This length scale is clearly less than the radius of the incoming ring but appears to be very similiar to the distribution of the radii of curvature of the deformed ring (shown in the inset of Fig.\,\ref{fig_prob}).

The mechanism for the production of the small rings is essentially that proposed by Svistunov\cite{svistunov95}. When the incoming ring reconnects with the deformed ring, the ring becomes even more distorted. Some of the excitations then have large enough amplitudes to produce a self-reconnection. There are occasionally up to three small rings emitted from different parts of the deformed ring. 

\section{Comparison with experiment}

The primary motivation of this paper was to seek to understand why small rings are created with high probability in experiments on interacting unidirectional vortex rings in superfluid $^4$He at very low temperatures\cite{walmsley14}. The experiments consisted of an injected pulse of charged rings with near identical radii each tagged by one electron trapped on the vortex core. Collisions could be detected due to changes in the shape of the pulse of collected electrons with shorter times of flight indicating the creation of small rings. The radii of the small rings would need to be less than half the radii of the pre-interaction rings to make a discernible difference to the time of flight.

Firstly, if we consider the interactions between circular rings (a primary interaction) then the impact parameter needs to be in the range $(b_1+b_2)/2 \lesssim b \lesssim b_2$ to produce small rings (where the radius of the smaller incoming ring is reduced by a factor of two or more). For rings of identical radii $(\Delta R\rightarrow 0)$ the mathematical expectation of a reconnection producing such a small ring is $\left(b_2^2-\left(b_1+b_2\right)^2/4\right)/\left(b_3^2-b_1^2)\right)\approx 0.4$, although this decreases for increased values of $\Delta R$ due to the rapidly increasing $b_3$ (on the other hand this also increases the probability of a reconnection actually occurring).

Secondly, the results from the previous section shows that the collision of rings with slow-moving large rings left behind from a previous merger (a secondary interaction), will produce small rings that also has a probability of $\simeq 0.4$ per reconnection (compared to a much smaller probability of only 0.07 for the equivalent case of circular rings with the same large value of $\Delta R$). It seems that if, initially, the beam of rings all have almost identical radii, then there needs to be a  primary interaction (two identical rings merging) to create a slow deformed ring, but once this has happened the majority of small rings in the experiment are most likely created though secondary collisions as an avalanche like pile-up will occur. The stream of small rings can also catch-up with and interact with any larger rings in front of them, with a broader range of impact parameters producing reconnections due to the large $\Delta R$.

A third scenario that we have not considered here, is that very small vortex rings are created in the vicinity of each reconnection. Several works have shown that the reconnection between anti-parallel lines can create a vortex ring cascade\cite{svistunov95,kerr11,kursa11}. The dissipative reconnection algorithm used in this paper tends to prevent this from being observed, although we note that high resolution simulations of reconnecting rings\cite{hanninen13,hanninen15} found no evidence of this process and it remains an open question as to whether this cascade can occur for realistic vortex configurations.

Our observation that small rings are created with high probability due to reconnections in the presence of strong deformations reinforces the view that the emission of small rings is likely to be a common feature in turbulent vortex tangles at low temperatures\cite{svistunov95,kozik09,kond12,nem14,laurie15}. The time-dependent vortex line density ${\cal L}(t)$ during the free decay of uncorrelated tangles (created through vortex ring collisions)\cite{walmsley08,zmeev15} in the zero-temperature limit was observed to be 
\begin{equation}
{\cal L} =1.2\zeta^{-1}\kappa^{-1} t^{-1},
\label{Lt}
\end{equation}
where $\zeta\approx 0.1$. These observations were in good agreement with values for $\zeta$ obtained from numerical simulations\cite{tsubota00,kond14}. In case the Kelvin-wave cascade of energy to smaller length scales is relatively inefficient, one can speculate that in these tangles the amplitude of Kelvin waves is kept on the verge of self-reconnections with  $A/R_0 \sim 0.5$ (here $R_0\sim \ell = {\cal L}^{-1/2}$ is the typical smoothed radius of curvature of vortex lines), i.\,e. a self-organized critical state. Hence, every new reconnection, that would further increase this amplitude, has a high probability of provoking one or more self-reconnections resulting in the emission of vortex rings of size $R' \ll \ell$. These small rings (with mean free path $\sim\ell^2/R')$ could then escape to the boundary, thus providing a channel for the transport of energy \cite{barenghi02,nem10,kond12}.  So, if the reconnnections occur at the rate, per unit volume, $\chi\kappa {\cal L}^{5/2}$ where $\chi \sim 0.1$,\cite{tsubota00} and each one effectively results in the pinching-off of a vortex ring of radius $R'$ (removing energy $\frac{(\Lambda_0-1)}{2}\rho\kappa^2 R'$, from Eq.\,\ref{E0}), the rate of energy removal becomes $\chi\frac{(\Lambda_0-1)}{2}\rho\kappa^3{\cal L}^{5/2} R'$. This should be compared with the measured rate of energy removal from the tangle \cite{walmsley08,zmeev15}, $\zeta\rho \kappa^3 {\cal L}^2$. Hence, for the pinching-off of small vortex rings to be the dominant mechanism of energy loss in quantum turbulence, the average size of small rings should be at least $R'/\ell = \frac{2\zeta}{\chi (\Lambda_0-1)} \sim 0.2$ -- which does not seem unreasonable.

\section{Summary}

We have used the vortex filament model to investigate the interactions between pairs of unidirectional vortex rings with a variable impact parameter. It was found that rings that have very similar initial radii, $\Delta R \ll R_{\mathrm{m}}$, are unlikely to reconnect; this is because the small relative velocity of the rings leads to there being sufficient time for non-local effects to push the rings sideways and away from each other. For those that do reconnect, the ratio of the mathematical expectation of a merger into a single ring to that for two daughter rings,  $\frac{b_3^2-b_2^2}{b_2^2-b_1^2} \approx 3:1$, favors a single large daughter ring. In contrast, when the difference in initial radii is large (and thus the relative velocity is high and the interaction time is short), the range of impact parameters where reconnections occur is in good agreement with a simple geometric model of straight trajectories of undeformed circular rings (that would have been predicted by the LIA without any non-local corrections); hence, the number of reconnection events resulting in either one or more daughter rings are nearly equal, $\frac{b_3^2-b_2^2}{b_2^2-b_1^2} \approx 1:1$. 

We have also considered the case where the larger ring is replaced with a deformed ring. The main difference is that the excess line length leads to more frequent self-reconnections with the result that the emission of small rings occurs with relatively high probability. It is likely that this mechanism lies behind recent experimental observations of small vortex rings created in a beam of unidirectional rings.

\begin{acknowledgments}
We thank Hongchao Xie for assistance in the early stages of this project and the anonymous referees whose comments helped improve the paper. This work was funded by the Engineering and Physical Sciences Research Council (grant no. EP/I003738). We acknowledge useful discussions and support from Manchester Research IT Services.  
\end{acknowledgments}

\end{document}